\documentclass[sigconf]{acmart}

\usepackage{multirow}
\PassOptionsToPackage{table}{xcolor}
\AtBeginDocument{%
  \providecommand\BibTeX{{%
    \normalfont B\kern-0.5em{\scshape i\kern-0.25em b}\kern-0.8em\TeX}}}

\copyrightyear{2020}
\acmYear{2020}
\setcopyright{acmcopyright}\acmConference[ATQAM/MAST'20]{Joint Workshop on Aesthetic and Technical Quality Assessment of Multimedia and Media Analytics for Societal Trends}{October 12--16, 2020}{Seattle, WA, USA}
\acmBooktitle{Joint Workshop on Aesthetic and Technical Quality Assessment of Multimedia and Media Analytics for Societal Trends (ATQAM/MAST'20), October 12--16, 2020, Seattle, WA, USA}
\acmPrice{15.00}
\acmDOI{10.1145/3423268.3423584}
\acmISBN{978-1-4503-8154-3/20/10}

\begin{document}

\title{Exploring Speech Cues in Web-mined \\ COVID-19 Conversational Vlogs}


\author{Kexin Feng}
\email{kexin@tamu.edu}
\affiliation{%
  \institution{Department of Computer Science and Engineering \\ Texas A\&M University}
  \city{College Station}
  \state{Texas}
}

\author{Preeti Zanwar}
\email{pzanwar@tamu.edu}
\affiliation{%
  \institution{Department of Epidemiology \& Biostatistics \\ Texas A\&M University}
  \city{College Station}
  \state{Texas}
}

\author{Amir H. Behzadan}
\email{abehzadan@tamu.edu}
\affiliation{%
  \institution{Department of Construction Science \\ Texas A\&M University}
  \city{College Station}
  \state{Texas}
}

\author{Theodora Chaspari}
\email{chaspari@tamu.edu}
\affiliation{%
  \institution{Department of Computer Science and Engineering \\ Texas A\&M University}
  \city{College Station}
  \state{Texas}
}


\begin{abstract}
The COVID-19 pandemic caused by the novel SARS-Coronavirus-2 (n-SARS-CoV-2) has impacted people's lives in unprecedented ways. During the time of the pandemic, social vloggers have used social media to actively share their opinions or experiences in quarantine. This paper collected videos from YouTube to track emotional responses in conversational vlogs and their potential associations with events related to the pandemic. In particular, vlogs uploaded from locations in New York City were analyzed given that this was one of the first epicenters of the pandemic in the United States. We observed some common patterns in vloggers' acoustic and linguistic features across the time span of the quarantine, which is indicative of changes in emotional reactivity. Additionally, we investigated fluctuations of acoustic and linguistic patterns in relation to COVID-19 events in the New York area (e.g. the number of daily new cases, number of deaths, and extension of stay-at-home order and state of emergency). Our results indicate that acoustic features, such as zero-crossing-rate, jitter, and shimmer, can be valuable for analyzing emotional reactivity in social media videos. Our findings further indicate that some of the peaks of the acoustic and linguistic indices align with COVID-19 events, such as the peak in the number of deaths and emergency declaration.
\end{abstract}




\maketitle

\section{Introduction}
The COVID-19 disease caused by the the novel SARS-Coronavirus-2 (n-SARS-CoV-2) became a pandemic in late 2019, and have infected tens of millions people around the world \cite{johns_hopkins}. The exact date of first report of the COVID-19 disease and the origins of the n-SARS-CoV-2 are undergoing scientific investigation.
Throughout this pandemic, governments have been encouraging people to stay home and reduce physical contact with others. This prolonged confinement and absence of face-to-face interaction in combination with negative feelings of anxiety caused by the pandemic s expected to result in significant emotional strain. Social media platforms, such as Weibo and Twitter, can potentially reveal individuals' emotional reactions to such impactful events and have been actively explored by researchers in social media analytics \cite{li2020impact,kleinberg2020measuring}.

In contrast to social networks and blogs, that rely mostly on written communication, conversational vlogs provide a valuable source of multimodal data for understanding subtle facets of emotion in communities and societies through the integration of spoken language and visual information. Conversational vlogs refer to a specific type of vlogging, in which whole or part of the shots depict a single person facing and talking to the camera \cite{biel2011vlogsense}. The richness of multimodal information presented in conversational vlogs can potentially provide a better understanding of the vloggers' attitude, feelings, and emotions compared to written text. Additionally, interactive cues in vlog videos, such as comments and number of upvotes or downvotes, can help identify how attitudes and emotions in the vlogs are propagated to the world. While few previous studies have conducted public sentiment analysis during a short period of the pandemic based on written text in social media~\cite{samuel2020covid}, to the best of our knowledge, multimodal analysis of conversational vlogs with the aim of a better understanding of public sentiments during the pandemic has not been examined.

To fill this gap, we collected 463 conversational vlogs from New York city, a major epicenter of the pandemic in the United States (U.S.). The examined vlogs uploaded between March 13, 2020 (date of announcement of the national emergency declared by the U.S. government) and June 1, 2020 \cite{dzhanova_2020,white_house}. We then applied a speech pre-processing pipeline to obtain the acoustic features indicative of prosodic changes on a weekly basis. We further analyzed the frequency of the words in the title and description of the YouTube videos to obtain a set of linguistic descriptors. Analysis of the acoustic and linguistic data found significant fluctuations across the span of the 11 weeks, some of which aligned with significant COVID-19 events, such as the peak in the number of deaths in New York City, as well as the stay-at-home order. Our pilot study provides preliminary insights into vloggers' emotional reactions during the period of the COVID-19 pandemic and contributes to better understating of emotion type and propagation in social media videos.

\section{Prior Work}
YouTube has been widely used by researchers in computing and social science, due to it is a great source of naturalistic and diverse real-life data \cite{susarla2012social,real2017youtube}. Vlogging became a social trend after 2010 \cite{gao2010vlogging}. Conversational vlogs are a specific type of vlogging, where usually an individual talking to a camera to share their ideas, views, or expertise about a topic of interest. This have rendered the content of vlogs a valuable source for researchers enabling the better understanding of people's behavior in social media \cite{gao2010vlogging}. Biel et al. used verbal and non-verbal cues of the vlogger to estimate the amount of social attention that the vlog would receive \cite{biel2011vlogsense}. Integrating vloggers' personality scores can further increase the accuracy of this task \cite{biel2011you}. Biel et al. further performed crowdsourcing experiments to investigate how vloggers were perceived by their audience \cite{biel2012good}. Researchers have also done sentiment analysis using comments posted under YouTube videos, and reviews posted for Movies. \cite{wollmer2013youtube,hajar2016using,savigny2017emotion,chen2017emotion}. 

Although previous studies have performed sentiment analysis in YouTube, conversational vlogs and emotion tracking of vloggers' experiences remains underexplored. This motivated us to explore the possible impact of social events on vloggers. To the best of our knowledge, our study is the first to investigate vloggers' emotions during the period of COVID-19 and their potential association with significant events of the pandemic. Our pilot study examines data collected by vloggers in New York, which was the first center of the pandemic in the U.S \cite{dzhanova_2020}. Findings from our work could provide a better understanding on how emotion is propagated in social media during during large-scale emergencies and life-changing events.

\section{Data}
\label{sec:data}
In this section, we discuss data collection and processing methods of conversational vlogs from YouTube videos.

\subsection{Data Description}
Given the recency of the COVID-19 pandemic and nonexistence of datasets of vlogs recorded during this period, we collected a new dataset from YouTube. Due to limitations of the YouTube API for collecting videos at a large-scale (e.g., many vlogs are from people who do not reside in the U.S., which is not related to our goal), we applied selenium webdriver, an automation testing tool to obtain YouTube videos \cite{avasarala2014selenium}. Selenium is able to simulate a human search behavior in a browser window on the YouTube website by scrolling down to the end of the search query while tracking the search results. In this way, the number of overseas videos is significantly decreased since the result of such traditional search method could possibly related to searcher's region. To maximize the number of retrieved videos, our queries included a variety of keywords through the combination of three components related to the event (e.g., COVID-19), behavior (e.g., vlog), and location (e.g., New York), as in Table \ref{tab:components}, resulting in a total of 18 combinations.

After removing duplicate videos from the search results, we obtained 4,265 videos potentially relevant to COVID-19 vlogs in New York City. We further collected video information, including title, description, duration, date published, number of views, and number of upvotes and downvotes, which could potentially become cues based on previous research \cite{biel2011vlogsense}. We then filtered out the videos before the March 13th, the date when U.S. national emergency is declared, resulting in a total of 3,021 videos. The end date is June 1st, the time when we performed the data collection. Among those videos, we manually examined each one to make sure they satisfy the following requirements:
\begin{itemize}
  \item Part of the video displays a conversational shot.
  \item The video has low or no background music or noise.
  \item The video is not recorded overseas.
\end{itemize}
This process resulted in a total number of 463 valid videos. Due to the subjectivity of this task, 13.24\% of the 3,021 videos (400 videos) were cross-examined by an additional annotator, yielding a Cohen's kappa coefficient of 0.703. The main reasons for the different labels between annotators are relying on annotator's subjective judgment about background music level and the ratio of conversational shots in each video. In order to obtain the videos within New York, we further selected the videos that included ``NY," ``NYC," or ``New York" in the title or corresponding description, which yielded a final number of 278 videos that were used in the rest of the analysis. 

\begin{table}[t]
\caption{Components relevant to keywords used in the Selenium tool for video mining.}
\begin{tabular}{c|c|c}
\hline
Event & Behavior & Location \\ \hline
quarantine & vlog & \multirow{3}{*}{\begin{tabular}[c]{@{}c@{}}New York\\ NY\end{tabular}} \\
covid-19 & vlogger &  \\
pandemic & vlogging &  \\ \hline
\end{tabular}
\label{tab:components}
\end{table}

\subsection{Speech Processing}
The audio data obtained from our study may contain multiple speakers, as well as non-speech segments corresponding to noise and music background. To address this challenge, we manually labeled a 5-second reference audio of the target speaker within each video, and did the speaker diarization by calculating the similarity between the reference audio and each target analysis window. Similarity metrics were calculated in a 256-dimensional d-vector space calculated by a deep learning model \cite{wan2018generalized}. A similarity score of 0.65 is used as a threshold and the window size is set to 125 milliseconds. By introducing the reference audio, this speaker diarization step can effectively identify non-speech segments or speech segments that do not belong to the target speaker.


\subsection{Acoustic Features}
Acoustic features focused on capturing prosodic changes, which are indicative of emotional information~\cite{schuller2009acoustic}. We extracted four prosodic features, namely loudness, zero-crossing-rate (ZCR), jitter, and shimmer, followed by four statistics descriptors of these features, that include mean, standard deviation, skewness, and the slope coefficient of the linear regression fit. For ZCR, we also extracted the min and max values since these two ZCR descriptors are used in other emotion-related tasks \cite{schuller2009interspeech}. To avoid the influence of possible extreme values, we first segmented the speech signal into 125ms windows, and subsequently obtained the statistics of the prosodic descriptors using a 25ms frame with a 10ms shift using OpenSmile toolbox \cite{eyben2010opensmile}. Finally, we calculated the average for each descriptor among all the windows. As a result, a 18-dimensional speech feature for each video was extracted.

\subsection{Linguistic Features}
As an additional source of information, we analyzed the words in the title and video description of each video and measured the word frequency. We further considered the words with the highest frequency from each video as linguistic measures.

\subsection{COVID Spread Data}
\begin{figure}[!htb]
  \centering
  \begin{minipage}{0.49\linewidth}
  \includegraphics[trim = 1cm 0cm 0cm 0cm, clip=true, scale=0.31]{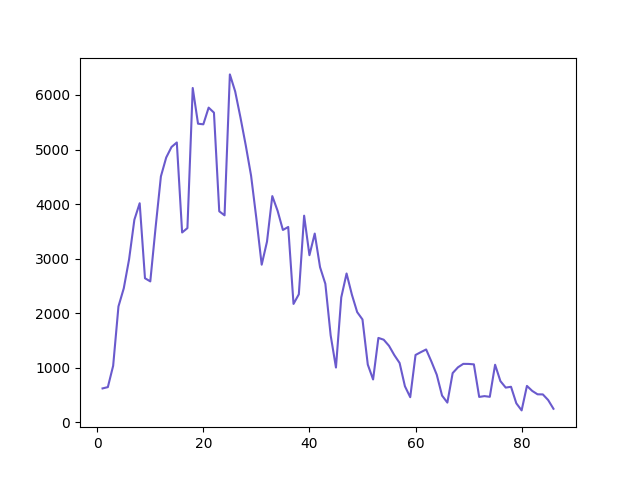}
  \centerline{\small{(a) {\it New COVID-19 cases}}}
  \end{minipage}
  \hfill
  \begin{minipage}{0.49\linewidth}
  \includegraphics[trim = 1cm 0cm 0cm 0cm, clip=true, scale=0.31]{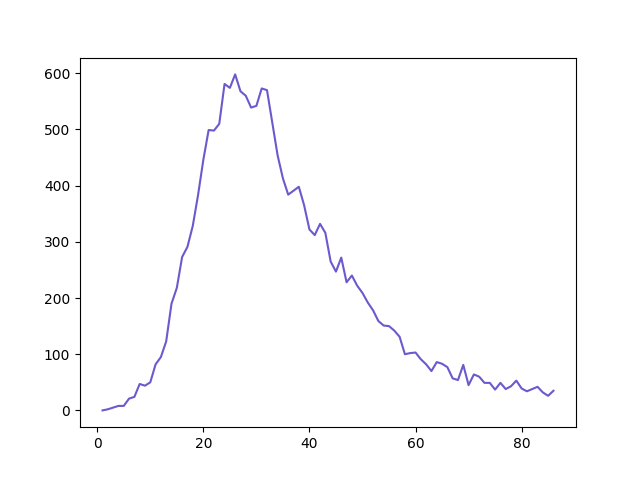}
  \centerline{(b) {\it New deaths from COVID-19}}
  \end{minipage}
  \vspace{-5pt}
 \caption{Visualization of number of daily new cases and deaths in New York City. The x-axis is the number of days from March 13th, while the y-axis indicates the number of people.}
\label{fig:gov data}
\end{figure}

To explore the connections between social media reactions and COVID-19 spread, we collected data related to COVID statistics. In particular, we used the NYC Open Data and measured daily number of new cases, new deaths, and hospitalized patients \cite{gov_data}. We only included the data after March 13rd to match the start of the search period for extracted videos, as visualized in Fig. \ref{fig:gov data}.

\begin{table*}[tb]
\caption{The starting date and number of videos within each group (week).}
\begin{tabular}{llllllllllll}
\hline
week number & 1 & 2 & 3 & 4 & 5 & 6 & 7 & 8 & 9 & 10 & 11 \\\hline
starting date & 3/13/20 & 3/20/20 & 3/27/20 & 4/3/20 & 4/10/20 & 4/17/20 & 4/24/20 & 5/3/20 & 5/11/20 & 5/18/20 & 5/26/20 \\
number of videos & 20 & 51 & 23 & 26 & 30 & 14 & 10 & 18 & 28 & 27 & 31 \\ \hline
\end{tabular}
\label{tab:date}
\end{table*}

\section{Data Analysis}
Our experiment aims to find if there is a potential connection between people's reactions in social media vlogs and the spread of COVID-19. Since there is generally a delay between video recording and posting online, we clustered the data into weekly bins, which is likely to address this delay.

\begin{figure*}[!htb]
  \centering
  \begin{minipage}{0.24\linewidth}
  \includegraphics[trim = 0cm 0cm 0cm 0cm, clip=true, scale=0.3]{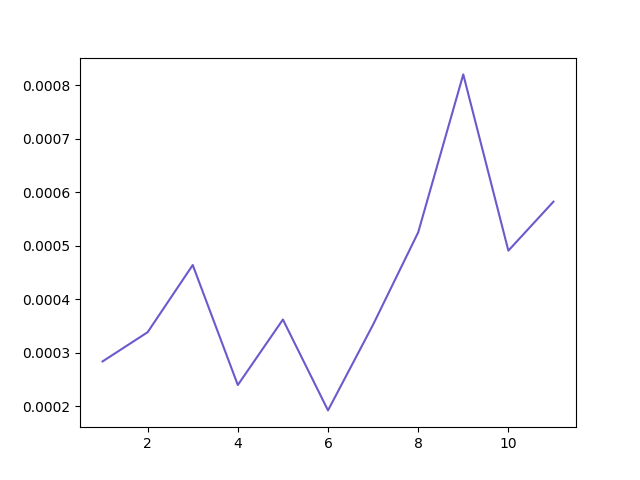}
  \centerline{(a) {\it Jitter linreg}}\medskip
  \end{minipage}
  \hfill
  \begin{minipage}{0.24\linewidth}
  \includegraphics[trim = 0.4cm 0cm 0cm 0cm, clip=true, scale=0.3]{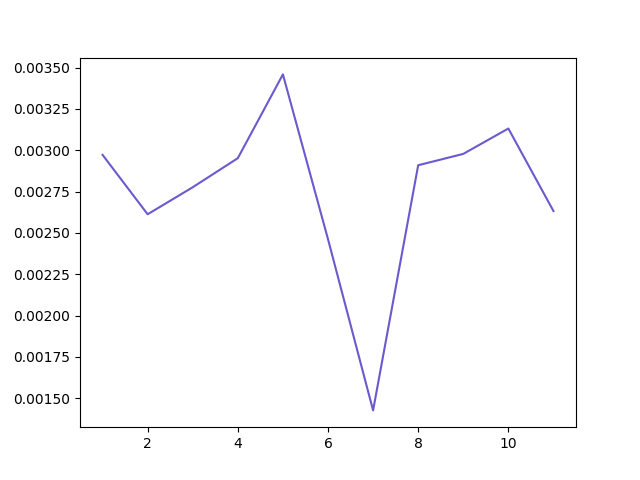}
  \centerline{(b) {\it Shimmer linreg}}\medskip
  \end{minipage}
    \begin{minipage}{0.24\linewidth}
  \includegraphics[trim = 0.4cm 0cm 0cm 0cm, clip=true, scale=0.3]{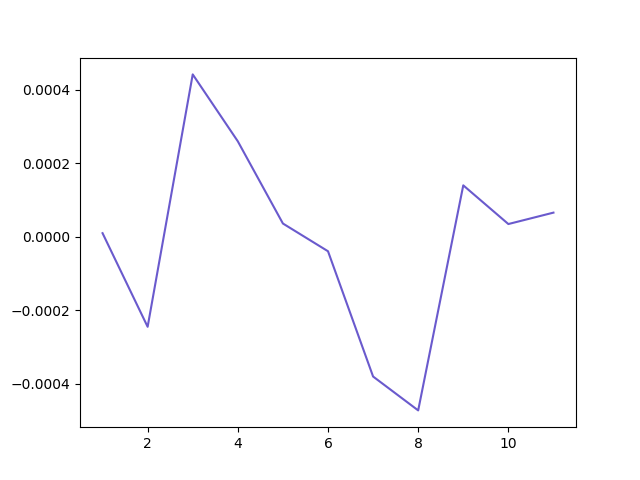}
  \centerline{(c) {\it ZCR linreg}}
  \end{minipage}
  \vspace{-5pt}
  \begin{minipage}{0.24\linewidth}
  \includegraphics[trim = 0cm 0cm 0cm 0cm, clip=true, scale=0.3]{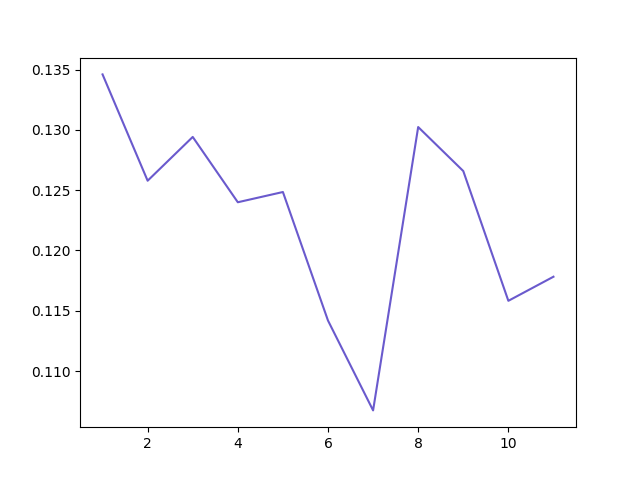}
  \centerline{(d) {\it ZCR mean}}
  \end{minipage}
  \vspace{-5pt}
 \caption{Visualization of jitter, shimmer, and zero-crossing rate (ZCR) extracted from videos. The x-axis is the number of weeks from initial video date (March 13th), while the y-axis indicates the value of this feature. Linreg refers to the slope of the linear regression to which the corresponding measure was fitted.}
\label{fig:peaks}
\end{figure*}

More specifically, starting from March 13th, we grouped videos published every 7 days, which resulted in 11 time periods ending on June 1st. The detailed start date information and number of videos in each time period can be found in Table \ref{tab:date}. For each week, we calculated the average of each acoustic feature to obtain a 18-dimensional weekly prosodic representation (Fig. \ref{fig:peaks}). The most frequent words and corresponding frequency of the title and video description over each week was also examined. We list the most frequent words within each week in Table \ref{tab:word freq}, and plot the largest frequency within target words (e.g., word 'quarantine' is the most frequent in most weeks) among weeks to explore potential connections with COVID-19 spread as shown in Fig. \ref{fig:word freq}.


\begin{table*}[tb]
\caption{Summary of most frequently used words and their corresponding frequency over each week.}
\resizebox{\textwidth}{!}{
\begin{tabular}{llllllllllll}
\hline
week number & 1 & 2 & 3 & 4 & 5 & 6 & 7 & 8 & 9 & 10 & 11 \\\hline
 & vlog 1.15 & new 1.12 & vlog 1.13 & vlog 1.12 & vlog 1.57 & vlog 1.43 & new 1.5 & vlog 1.44 & new 2.11 & show 1.07 & vlog 1.06 \\
 & show 0.9 & vlog 1.08 & \textit{coronavirus 1.13} & \textit{quarantine 1.0} & \textit{quarantine 1.37} & show 1.14 & vlog 1.4 & \textit{quarantine 1.06} & life 2.04 & vlog 1.04 & show 1.06 \\
 & new 0.8 & show 1.02 & show 1.0 & show 1.0 & new 1.167 & nyc 1.0 & york 1.3 & show 0.94 & day 1.75 & nyc 1.0 & new 0.87 \\
 & nyc 0.65 & \textit{quarantine 0.94} & new 1.0 & new 0.69 & show 1.07 & \textit{quarantine 0.93} & \textit{quarantine 0.9} & new 0.72 & york 1.75 & \textit{quarantine 0.81} & nyc 0.84 \\
\multirow{-5}{*}{\begin{tabular}[c]{@{}l@{}}most frequent\\ words and \\ their \\ frequency\end{tabular}} & \textit{coronavirus 0.6} & york 0.80 & nyc 0.87 & nyc 0.62 & york 0.9 & 19 0.64 & family 0.9 & nyc 0.61 & \textit{quarantine 1.25} & new 0.7 & \textit{quarantine 0.61} \\ \hline
\end{tabular}}
{\raggedright The COVID-19 related words are marked with \textit{italic}. \par}
\label{tab:word freq}
\end{table*}

\subsection{Results and Discussion}
After plotting the variation of the 18 acoustic features across all the weeks, common patterns seem to emerge in some of the extracted features. Particularly, we found that the slope of the linear regression fit is the most consistent descriptor compared with mean, skewness, or standard deviation, potentially due to its ability to capture temporal trends. As for the four acoustic features (loudness, ZCR, jitter, and shimmer), the loudness was not able to reveal meaningful patterns. A possible reason could be that the loudness is highly dependent on the recording conditions (e.g., microphone, distance from the microphone), making it highly variable across videos. Based on Fig. \ref{fig:peaks}, three peaks at week 3, 5, and 8-9 can be observed among the other three acoustic features.

\begin{figure}[!htb]
  \centering
  \includegraphics[trim = 1cm 0cm 0cm 0cm, clip=true, scale=0.3]{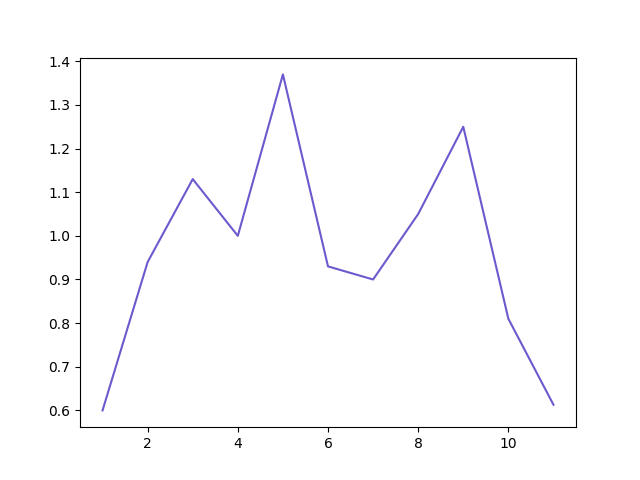}
 \caption{Visualization of largest frequency of target words among weeks. The x-axis is the number of week from March 13th, while the y-axis indicates the word frequency.}
\label{fig:word freq}
\end{figure}

Next, we explored the change of word frequency among those weeks. As observed in Fig. \ref{fig:word freq}, the frequency of negative words relevant to COVID-19 is higher in weeks 3, 5, and 8-9. In the remaining weeks, the frequency of such words is rarely greater than 1. The word frequency analysis is thus consistent with the change observed across acoustic features. 

Finally, we explored potential connections between the acoustic and linguistic trajectories and the COVID-19 spread. According to Fig. \ref{fig:gov data}, the number of daily new cases reached a peak in New York City around week 3 (day 20), while the number of daily new deaths reached to the peak around week 5 (day 35). During weeks 8 and 9, The Governor of New York extended the PAUSE order as well as the state of emergency for the New York state, which added to the quarantine period \cite{nbc}. Although we cannot draw direct comparisons, due to the fact that acoustic and linguistic measures might be confounded by multiple factors, we observed similar spikes in weeks 3, 5, and 8-9 in some of the acoustic data. These trajectory similarities might suggest that COVID-19 related events can influence vloggers' sentiments, but a more thorough analysis is needed to better understand the contextual factors causing such spikes in the acoustic and linguistic trends.


Even though our results indicate fluctuations in the acoustic and linguistic features which might be relevant to COVID-19 events, there are various limitations in our study. First, the data collection step heavily relies on the YouTube video search results. Possible bias could occur in this process, such as potentially more popular videos being retrieved first. Also, even though we took actions to make sure all the videos used in our study are from New York, we cannot exactly justify the video is from New York City or the state of New York. Finally, contextual factors that capture the content of each video have to be taken into account in a more thourough analysis.

\section{Conclusions}
In this paper, we explored the possibility of understanding public sentiments during the COVID-19 pandemic through multimodal content of social media. We selected New York City, because it was one of the first epicenters of COVID-19 pandemic. We collected our own dataset from YouTube and provided a complete pipeline to pre-processing and analyzing real-life audio data. We then extracted acoustic features and observed common patterns from three features (jitter, shimmer, and ZCR). This pattern was also consistent with a word frequency analysis performed on video title and description and can be potentially explained by taking into account the timing of major COVID-19 related events occurring in New York City during this time.

As part of our future work, we plan to extend our study to additional geographical locations, and explore the influence of gender, age, as well as other potential factors on viewers' reactions to social media content. Finally, we will analyze additional cues from the videos, such as facial expression and linguistic cues obtained from the vloggers' speech, to obtain a better understanding of social media videos.

\section{Acknowledgement}
This work was supported by the Texas A\&M Institute of Data Science (TAMIDS) through the Data Resource Development Program. The authors would like to thank Alexandria Curtis, Texas A\&M Computer Science \& Engineering student, for her help in annotating the conversational vlogs.

\bibliographystyle{ACM-Reference-Format}
\bibliography{refs}


\end{document}